\documentstyle[twocolumn,aps,prl]{revtex} 

\begin{document}             

\draft

\twocolumn[\hsize\textwidth\columnwidth\hsize\csname @twocolumnfalse\endcsname

\title{Charge transfer in molecular conductors --- oxidation or reduction?}
\bigskip

\author{Avik W. Ghosh, Ferdows Zahid and Supriyo Datta }

\address{School of Electrical and Computer Engineering,
Purdue University,
West Lafayette, IN 47907-1285, USA.} %

\author{Robert R. Birge}

\address{Dept. of Chemistry,
University of Connecticut,
Storrs CT 06269-3060, USA} %

\maketitle
 
\widetext
\begin{abstract}
We discuss the nature of charge transfer in molecular conductors upon
connecting to two metallic contacts and imposing a voltage bias across
them. The sign of the charge transfer (oxidation vs. reduction) depends
on the position of the metal Fermi energy with respect to the molecular
levels. In addition, the charge transfer depends on the strength of the
coupling (chemisorption vs. physisorption) with the contacts. A convenient 
way to establish the nature and onset of the charge transfer and the
corresponding features in the I-V is to draw an energy level diagram for
each spin species.  Starting from such a level diagram, we argue that
transport in the Tour-Reed switching molecules, which consist of a
central phenyl ring with a nitroamine redox center, involves the oxidation of 
a highest occupied molecular orbital (HOMO)-based level. 
\end{abstract} 
\bigskip

]
\narrowtext

\bigskip

In recent years, there have been many reports of conductance
characteristics measured for individual molecules or small ensembles of
oriented molecules \cite{r1}.
Understanding transport in molecular conductors is quite challenging
because of its special status between solid state physics and
molecular chemistry. On the one hand, the system differs from
conventional solid-state devices where conduction is understood in
terms of alignment of a band structure that is largely unaffected by 
charge transfer or the nature of the contacts. On
the other hand, in contrast to conventional chemistry, the molecule is
not an isolated unit in equilibrium; coupling with metallic
contacts makes the system an open one, capable of freely exchanging a
fractional amount of charge to and from the contacts. In addition in
the presence of a voltage bias across the molecule, the system is out of
equilibrium and equilibrium statistical mechanical techniques become
inapplicable. Predicting I-V characteristics of molecular conductors
involves therefore a suitable combination of techniques and insights
both from chemistry as well as from mesoscopic physics.

\begin{figure}
\vspace*{3.2in}
\includegraphics{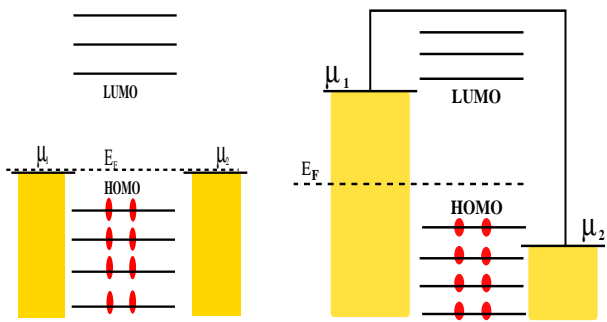}
\vskip -2cm
\caption{Simplified description of conduction in a mesoscopic device:
(a) {\it{Left:}} At equilibrium (V = 0) the metal contact
electrochemical potential is given by $\mu_{1,2} = E_F$. The molecular
levels are discrete, but end up being  broadened by coupling with
contacts. (b) {\it{Right:}} Under bias (V$ > 0$) the system is out of
equilibrium as the contact potentials split, $\mu_{1,2} = E_F \mp
eV/2$. Only the levels lying between $\mu_1$ and $\mu_2$ are filled
from one side (oxidized) and emptied from the other (reduced) with a
resultant conduction of current.  Levels lying below or above both
electrochemical potentials stay filled or empty and do not contribute
to current at zero temperature.} \label{f1} \end{figure}

We have recently developed a self-consistent, ab-initio procedure for
calculating I-V characteristics of molecular conductors \cite{r2}. In
this procedure, we self-consistently combine the outputs of a standard quantum chemical
software (GAUSSIAN98) with a non-equilibrium Green's function (NEGF)
formalism for transport in an open, voltage-biased system. The method
takes into account the hybridization of molecular levels with the contacts exactly,
and can be used both for weak and strong contacts. In this paper, we
will focus on an intuitive picture that can be used to gain qualitative
insight into the nature of conduction through a molecule. This picture
is particularly useful when the contacts are weakly coupled, and issues
related to level broadening are minor.  The first step is to obtain the
energy levels of an isolated molecule and locate the Fermi energy $E_F$
of the metallic contacts, as shown in Fig.~\ref{f1} (a).  Under bias
the electrochemical potentials $\mu_{1,2}$ of the contacts separate as
shown in Fig.~\ref{f1}(b). We wish to address the following questions
in this paper: (a) which molecular levels are involved in the
conduction process? (b) Does the molecule get oxidized ($M
\rightarrow M^+$) or reduced ($M \rightarrow M^-$) at steady state?

{\it{Equilibrium energy level diagram ($V = 0$)}}. 
We start by obtaining an equilibrium energy level diagram as in Fig.~\ref{f1}(a).
The energy levels for a given molecule are obtained using a standard quantum chemical 
software like Hyperchem, MOPAC or GAUSSIAN98. The equilibrium  Fermi energy 
of gold is known to be $E_F = - 5.1$ eV. It is important
to note that semi-empirical programs like Extended H\"uckel Theory do not give the
correct absolute value for molecular energy levels. In that case, the energy
levels for gold must also be computed using the same program for consistency 
\cite{rHall}.

{\it{Which levels conduct?}}.
Fig.~\ref{f1} shows a schematic description of transport through a
mesoscopic device. At equilibrium (Fig.~\ref{f1}(a)) the metal contacts
impose their common electrochemical potential on the molecule, equal to
the metal fermi energy $E_F$. Coupling of the molecule with the
contacts broadens each highest occupied (HOMO)/lowest unoccupied (LUMO) molecular
orbital-based level, so that the number of electrons
below the fermi energy is now fractional (the deviation of this number
from the number of electrons for the isolated molecule is a measure of
charge transfer between the molecule and the contacts at equilibrium). When a voltage
bias is imposed (Fig.~\ref{f1}(b)), the system is driven out of
equilibrium and the contact potential splits: $\mu_{1,2} = E_F \mp
eV/2$. (Note that in contrast to our earlier convention of fixing energy
levels and varying the contact potentials with bias by different
amounts depending on the ratio of couplings \cite{r4,r5}, we now vary
the electrochemical potentials as above and let the energy levels
adjust accordingly). For adiabatic, reflectionless contacts and
ballistic transport, the contact potentials are imposed separately on
the right and left-moving electrons \cite{r3}. As long as the two
electrochemical potentials lie in the gap, all HOMO levels lie below
both potentials and are filled at zero temperature, the LUMO levels lie 
above the potentials and are empty, and there is no current. When the bias is
sufficient that a level is crossed by one of the electrochemical
potentials,  that  level is emptied out (oxidized) from one side but
filled (reduced) from the other, with a resultant onset of current.
Thus only levels which lie between the two electrochemical potentials
contribute to conduction at zero temperature (at finite temperature,
this range is extended by the thermal energy $\pm k_BT$). The onset of current is
determined by the proximity of the Fermi energy to the molecular level
closest to it.

{\it{Oxidized or reduced?}}.
The coupling to the contacts of an extended molecule is described by a
non-hermitian self-energy matrix $\Sigma$ calculated in an appropriately 
chosen basis \cite{r3}. This self-energy effectively partitions the problem 
into the molecular part and the contact, and leads to a conceptual 
simplification of the problem. The broadenings of the levels (inverse
lifetime) are described by the anti-hermitian components $\Gamma_{1,2}$
of $\Sigma_{1,2}$.  For fermi functions $f_{1,2}$ corresponding to the
two contact potentials $\mu_{1,2}$, the steady-state occupancy of a
molecular level is given by $f=(f_1\Gamma_1 + f_2\Gamma_2)/(\Gamma_1 +
\Gamma_2)$. If the level lies between $\mu_{1,2}$ at zero temperature,
$f_1=0$ and $f_2 = 1$ so that the occupancy is given by $f =
\Gamma_2/(\Gamma_1 + \Gamma_2)$. One of the contacts is trying to add electrons
(reduce), while the other is trying to remove them (oxidize). If the left contact 
is stronger than the right contact ($\Gamma_1 \gg \Gamma_2$), the level is emptied
(oxidized) while in the opposite case the level stays filled (reduced).
In either case, a current flows due to the competition between
oxidation and reduction.  Typically for a self-assembled monolayer, one 
end of the molecule (usually consisting of a thiol group chemisorbed on a gold 
surface) has a strong contact, while the other end is physisorbed and 
has a weak contact. This means that for a positive bias on the physisorbed 
end, the molecule remains neutral, while for opposite bias, charging occurs 
from the strongly contacted chemisorbed end \cite{rPRL}.

For a high work function metal, $E_F$ usually lies near the HOMO
level. This is an issue that clearly needs a lot more attention, since the
precise location of $E_F$ depends on the model for the contacts, as well as
the method of calculating the molecular Hamiltonian. A third terminal (gate) \cite{rSchon}
can help resolve some of these issues. One of the consequences of a self-consistent evaluation of
energy levels is that as we raise the fermi energy, the levels float up
by an amount  equal to the charging energy (for a solid, this is
roughly given by the capacitance of a sphere of radius equal to the
size of the electronic wave function). For a molecule with strong
coupling to at least one contact, or for a molecule in solution, the
effective electronic size is infinite, so that the charging energy is
small, and classical STM theory holds. For a weakly coupled system,
charging effects lead to the levels following the Fermi energy, so that
the conduction tends to remain HOMO-based. We verified this for the
phenyl-dithiol molecule (PDT) where we couple the outputs of a GAUSSIAN98 \cite{r6} evaluation of energy levels using a Becke-3 parameter
exchange and Perdew-Wang 91 correlation (B3PW91) approximation
\cite{r7,r8}, with a non-equilibrium Green's function (NEGF) evaluation
of the density matrix \cite{r4,r5}, and iterate to convergence
self-consistently \cite{r2}.  The basis set used was the Los Alamos National Lab
set for effective core potentials of the double $\zeta$ type (LANL2DZ)
\cite{r9,r10}.  The converged energy levels increase with increasing
$E_F$ at the rate of $~ 4$eV/Volt, which corresponds to the
capacitative charging energy of a sphere of size $~5 \AA$.

\begin{figure}
\vspace*{3.2in}
\includegraphics{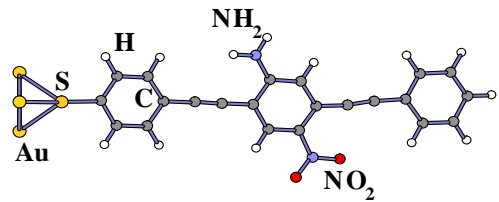}
\vskip 6cm
\hskip -1.0cm\includegraphics{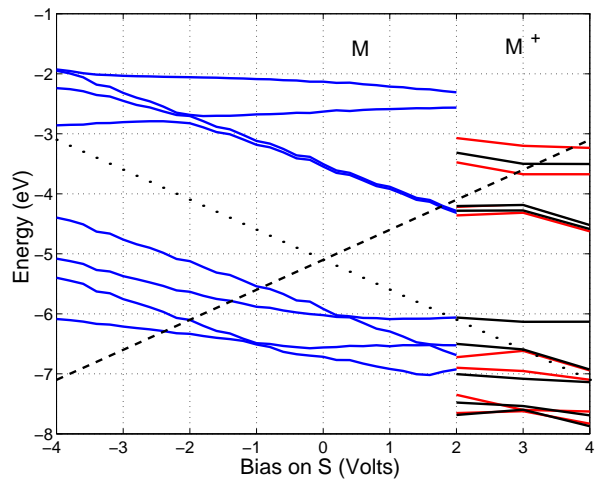}
\vskip -1.5cm
\caption{{\it{Top:}} Tour-Reed molecule with nitroamine side-group,
extended to include three gold atoms from the contact surface bonded
directly to sulphur. The molecule has a strong chemisorbed bond at the 
sulphur side and a weak physisorbed one at the other.
{\it{Bottom:}} Energy levels for the neutral molecule ($M$) as a
function of voltage applied to the sulphur end.
 The levels Stark shift with applied voltage. There are some gold-based levels
running roughly parallel to the electrochemical potential on sulphur $\mu_1$ (dotted line), that do not contribute to
conduction. The Fermi energy $E_F = -5.1$ eV is closer to the HOMO
levels than the LUMO ones (ignoring the trivial gold-based levels). At
positive bias, the HOMO level is crossed first with the strong sulphur
contact (dotted line) around 2 V, which empties the level out (oxidation).
Subsequently, we draw the energy levels for the oxidized species $M^+$,
which are spin nondegenerate (red and black lines).  At negative bias, the HOMO levels are
crossed with the physisorbed carbon contact (dashed line), which does not 
charge up the system due to the weakness of the contact coupling. The 
resulting I-V is expected to be strongly asymmetric, and exhibits charging 
effects arising from an oxidation step at around 2 V.} 
\label{f3} 
\end{figure}

{\it{Level diagram under bias ($V \neq 0$)}}.
As an illustrative example, we have plotted the energy levels of the
Tour-Reed molecule 2'-amino-
4,4'-di[ethynyl]phenyl-5'-nitro-1-[thioacetyl]benzene containing the
nitroamine redox center \cite{r11}, as a function of voltage applied to
the sulphur end (Fig.~\ref{f3}). We consider an extended molecule
including three surface gold atoms bonded to sulphur and obtain energy
levels analogous to Fig.~\ref{f1}, including the effect of
voltage-induced Stark shifts of the molecular orbitals.  Exchange and
correlation effects are included by performing a B3PW91 calculation
with a LANL2DZ basis set using the GAUSSIAN 98 software. In the
Tour-Reed experiment, one end of the molecule is self-assembled onto
gold using sulphur ``alligator clips'', leading to a strongly coupled
chemisorbed contact, while the other end is a carbon atom weakly
coupled to an evaporated gold contact. Charging occurs when the
electrochemical potential $\mu_1$ of the sulphur end (dotted line in
Fig.~\ref{f3}) crosses a molecular level. From Fig.~\ref{f3}, we see
that for positive bias on sulphur,  we first cross a HOMO level by the
stronger sulphur contact, which oxidizes the molecule (the first LUMO
level is actually a nonconducting gold-based level, as we discuss
below). For negative bias, the
HOMO level is crossed by the weaker carbon contact, so the level stays
filled and there is no charging. This means that the I-V should be
strongly asymmetric, as is indeed observed. \footnote{For a spatially
symmetric molecule, such contact-induced charging asymmetries alone 
determine the asymmetry in the I-V \cite{rasymm}. However for spatially 
asymmetric molecules such as the Tour-Reed example here, the I-V asymmetry 
is much larger, and given predominantly by the unequal Stark shifting of 
the intrinsic molecular wavefunctions (the asymmetric Stark shifting is
seen in Fig.~\ref{f3}).} The sulphur contact
potential $\mu_1$ crosses the HOMO level at around 2 V, whereupon the
molecule is oxidized from $M$ to $M^+$. This is important, since the
precise nature of the charging process is an important question raised
in the literature \cite{r12,r13}. Although cyclic voltammetric curves
seem to suggest reduction processes as being operative, one has to
remember that in such measurements counterions are present in solution,
and substantially compromise any charging effects that may dominate in
solid state.

When $M$ changes to $M^+$, we need a whole new set of energy levels
(Fig.~\ref{f3}).  Two changes happen: (i) since one electron of a
particular spin is pulled out by the contact, the spin degeneracy is
removed; (ii) secondly, the removal of one electron of given spin leads
to charging effects which cause all the levels of $M^+$ obtained from
GAUSSIAN98 to float down by about 2 volts.  This is unphysical, because
GAUSSIAN98 models an isolated molecule with image charges at infinite
distance. In the presence of image charges on the contacts, the charging
energy will be lower. We therefore raise the energy levels of $M^+$ so as to
align the energy level of the semi-occupied molecular orbital (SOMO) with the
HOMO of the neutral species ($M$). Such an alignment is motivated
by our assumption that in the absence of correlations and molecular
reorganization, the ionization potential of $M$ equals the electron
affinity of $M^+$. This assumption is a point that requires further work, and 
should come automatically out of a proper ab-initio
theory that includes the effect of image charges on the contacts. 
Note that in order to obtain spin-dependent charging effects
that shift all levels except the SOMO substantially, one needs to do
a spin-unrestricted (UHF/LSDA) calculation that breaks the spin degeneracy 
of the energy levels. 

It is interesting to note that for an extended molecule, there are some
gold-based levels. These can be identified either from their wave functions
that are localized strongly on gold, or from their energy levels; the levels
run roughly parallel to the electrochemical potential $\mu_1$ of the contact on the
chemisorbed side (dotted line in Fig.~\ref{f3}). For small molecules these 
gold levels, as well as metal-induced gap states (MIGS) that arise when a
self-energy is included in our calculation, end up contributing to the 
conductance through direct tunneling processes. For large molecules such 
states hybridize with molecular wave functions, but don't contribute directly 
to tunneling, so they can be ignored.  However the gold levels can still 
influence the positions of the other levels through level-repulsion, so the 
contacts affect the voltage at which a HOMO level is crossed and oxidation 
occurs. 

We have shown that in self-assembled monolayers such as the Tour-Reed
molecules containing the nitroamine redox center, charge transfer
occurs due to oxidation from the strongly contacted end involving the
level closest to the Fermi energy. We are currently investigating the
mechanism of switching thereafter.  Understanding the behavior of
a molecular switch is important from the point of view of molecular
electronics. Several switching molecules have been reported
\cite{r11,r14}, and various possible mechanisms suggested
\cite{r13,r15}.  A charged species could deform in an applied field,
and the electromechanical effect could switch off the current.  If the
added charge is localized on a sidechain of the molecule, it could rotate
the central benzene ring attached to it in the 
presence of the field, distortiing the conjugated electron wave
function enough to switch current off. A more complicated cooperative
effect involving the whole self-assembled monolayer could also arguably
lead to a current switch.  Prior to a full self-consistent ab-initio
calculation of transport properties including contact effects, it is
important to develop an intuitive understanding of transport in the
isolated species. An energy level diagram as illustrated above can
serve as a quick intuitive tool to answer some basic questions such as
the onset of current, whether an electron is being added or removed, or
the existence of an asymmetry in the I-V.

We would like to thank Prashant Damle and Deepak Singh for useful
discussions. This work was supported in part by the Defense Advanced
Research Projects Agency (DARPA) and the US Army Research Office (ARO)
under grants number DAAH04-96-1-0437 and DAAD19-99-1-0198.

\end{document}